\documentclass[twocolumn,showpacs,showkeys,preprintnumbers,amsmath,amssymb]{revtex4}

\usepackage{graphicx}
\usepackage{dcolumn}
\usepackage{bm}

\begin{document}
\title{Separated Oscillatory Fields for High-Precision 
Penning Trap Mass Spectrometry\\}

\author{S.~George$^{1,2}$}
\email[Correspondence to: ]{george@uni-mainz.de}
\author{S.~Baruah$^{3}$}
\author{B.~Blank$^{4}$}
\author{K.~Blaum$^{1,2}$}
\author{M.~Breitenfeldt$^{3}$}
\author{U.~Hager$^{5}$}
\author{F.~Herfurth$^{1}$}
\author{A.~Herlert$^{6}$}
\author{A.~Kellerbauer$^{7}$}
\author{H.--J.~Kluge$^{1,8}$}
\author{M.~Kretzschmar$^{2}$}
\author{D.~Lunney$^{9}$}
\author{R.~Savreux$^{1}$}
\author{S.~Schwarz$^{10}$}
\author{L.~Schweikhard$^{3}$}
\author{C.~Yazidjian$^{1}$}

\affiliation{$^{1}$GSI, Planckstra{\ss}e 1, 64291 Darmstadt,
Germany} 
\affiliation{$^{2}$Johannes Gutenberg-Universit\"at,
Institut f\"ur Physik, 55099 Mainz, Germany}
\affiliation{$^{3}$Ernst-Moritz-Arndt-Universit\"at, Institut f\"ur
Physik, 17487 Greifswald, Germany} 
\affiliation{$^{4}$Centre
d'Etudes Nucl$\acute{\mbox{e}}$aires de Bordeaux-Gradignan, 33175
Gradignan Cedex, France} 
\affiliation{$^{5}$University of
Jyv\"askyl\"a, Department of Physics, P.O. Box 35 (YFL), 40014
Jyv\"askyl\"a, Finland} 
\affiliation{$^{6}$CERN, Division EP, 1211
Geneva 23, Switzerland} 
\affiliation{$^{7}$Max-Planck Institut f\"ur
Kernphysik, 69117 Heidelberg, Germany}
\affiliation{$^{8}$Ruprecht-Karls-Universit\"at, 
Physikalisches Institut, 69120 Heidelberg, Germany}
\affiliation{$^{9}$CSNSM-IN2P3-CNRS, 91405 Orsay-Campus, France}
\affiliation{$^{10}$NSCL, Michigan State University, East Lansing,
MI 48824-1321, USA}

\date{\today}

\begin{abstract}
Ramsey's method of separated oscillatory fields is applied
to the excitation of the cyclotron motion of short-lived ions in a
Penning trap to improve the precision of their measured mass.  The
theoretical description of the extracted ion-cyclotron-resonance
line shape is derived out and  its correctness demonstrated 
experimentally by measuring the mass of the short-lived $^{38}$Ca 
nuclide with an uncertainty of $1.6\cdot 10^{-8}$ using the ISOLTRAP
Penning trap mass spectrometer at CERN. The mass value of the 
superallowed beta-emitter $^{38}$Ca is an important contribution 
for testing the conserved-vector-current hypothesis of the electroweak 
interaction. It is shown that the Ramsey method applied to mass 
measurements yields a statistical uncertainty similar to that obtained 
by the conventional technique ten times faster.
\end{abstract}
                          
\maketitle

In 1989 the Nobel prize in  physics was awarded to N.F. 
Ramsey~\cite{Rams1990} in recognition of his molecular beam resonance
method with spatially separated oscillatory fields proposed 40
years earlier~\cite{Rams1949,Rams1950}. In 1992 G. Bollen 
\emph{et al.}~\cite{Boll1992} demonstrated the use of 
time-separated oscillatory fields for the excitation of 
the cyclotron motion of an ion confined in the Penning trap 
mass spectrometer ISOLTRAP.  This and experiments performed later 
at SMILETRAP~\cite{Berg2002} showed that the method may 
improve the precision of mass measurements with Penning 
traps - on the condition that the shape of the detected 
ion-cyclotron-resonance could be put on a sound theoretical basis.
In the case of exotic nuclides, the extremely low production 
rates make precision measurements challenging because of the
time required to accumulate the necessary statistics.  The Ramsey
method would therefore allow accumulation of sufficient statistics
in a much shorter time, which is of prime importance for on-line measurements.

In this Letter, we introduce the correct theoretical description of
the application of the Ramsey method to ions stored in a Penning 
trap. We also demonstrate its validity for the first time with 
an on-line mass measurement of $^{38}$Ca ($T_{1/2}$ = 440(8) ms). 
This short-lived nuclide is used for testing the 
conserved-vector-current hypothesis (CVC) of the Standard 
Model of fundamental interactions, which postulates a vector-current 
part of the weak interaction unaffected by the strong 
interaction. Thus, the decay strength of all superallowed 
$0^+\rightarrow0^+$ $\beta$-decays is independent of the 
nuclei except for theoretical corrections~\cite{Hard2005a,Hard2005b}.
Here, a relative mass precision of $10^{-8}$ is required.

Comparisons with the conventional excitation 
scheme~\cite{Grae1980,Koen1995} show that the line width of the 
resonance is reduced by almost a factor of two and the statistical 
uncertainty of the extracted resonance frequency is more than a 
factor of three smaller. It is shown that the
Ramsey method allows a measurement with the same statistical
uncertainty, but ten times faster. This is an enormous improvement 
in the very active field of on-line high-precision mass measurements 
with Penning traps at radioactive beam facilities~\cite{Blau2006}.

The prerequisite for the successful implementation of the Ramsey 
method to stored ions is the detailed understanding of the observed 
time-of-flight cyclotron resonance curves using time separated 
oscillatory fields. While here only the most important parts of the 
theory and its experimental confirmation is given, a detailed 
presentation will be published elsewhere.
\begin{figure}
\resizebox{0.4\textwidth}{!}
{
 \includegraphics{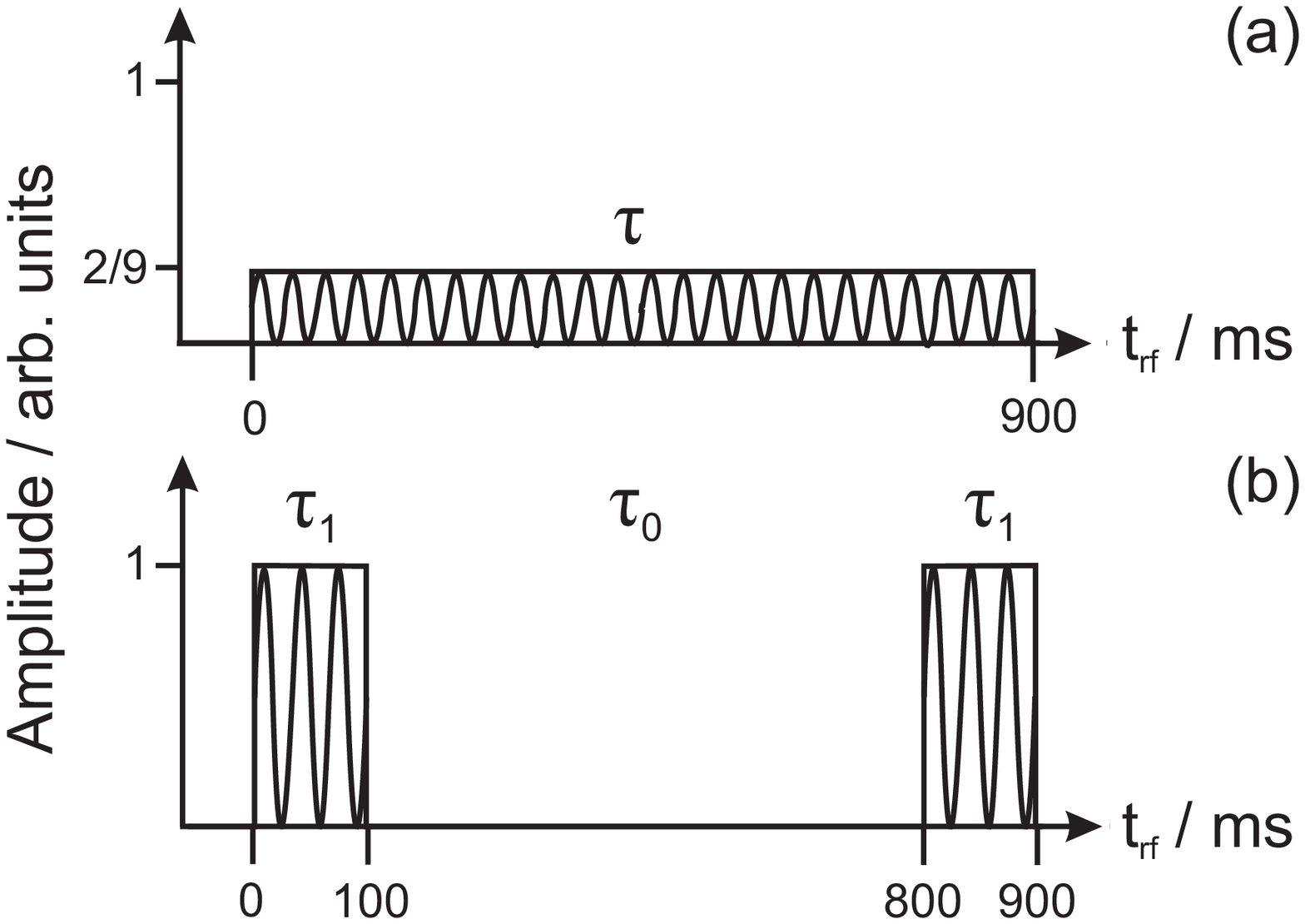}
   } \vspace*{0.0cm}
   \caption{Excitation schemes: (a) conventional excitation 
   with a continuous rf pulse (here $\tau$=900 ms), (b) 
   excitation with two 100-ms Ramsey pulses $\tau_1$ interrupted 
   by a 700 ms waiting period $\tau_0$.} \label{AnregSchema}
\end{figure}

In a Penning trap an ion with a charge-to-mass ratio q/m 
is stored in a strong homogeneous magnetic field $B_0$ 
combined with a weak electrostatic quadrupole
field. The mass measurement is performed via the determination 
of the cyclotron frequency $\nu_c=qB_0/(2\pi m)$. For the 
basic theory of the ion motion in Penning traps we refer 
to the review article of Brown and Gabrielse~\cite{Brow1986}. 
Our considerations here focus on the process of conversion of
the magnetron mode into the cyclotron mode under the influence of
the quadrupolar rf-field in the Penning trap, extending previous 
studies by one of us \cite{Kret1999} of this process in a
quantum mechanical framework. In this work the
effective interaction of a quadrupolar field with driving frequency
$\nu_d\approx\nu_c$ with a trapped ion is described in terms of the
annihilation ($a$) and creation ($a^{\dagger}$) operators of the
magnetron (-) and cyclotron (+) oscillators, the phase of the quadrupole 
field $\phi_{d}(t)=\omega_{d}t+\chi_{d}$
with $\omega_{d}=2\pi\nu_d$,  and a coupling constant $g$ that is
proportional to the amplitude of the quadrupolar field:
\begin{equation}
H_{1}(t)=\hbar g \left(e^{-i\phi_{d}(t)}a_{+}^{\dagger}
(t)a_{-}(t) + {\rm h.c.} \right), \label{int}
\end{equation}
\lq\lq h.c." denotes the Hermitian conjugate of the first term.
Addition of this interaction to the Hamiltonian of an ion in an 
ideal Penning trap yields a model system for which Heisenberg's 
equations of motion can be solved rigorously. 
The interaction (\ref{int}) has the important property that the total number
of quanta in the magnetron and cyclotron oscillators is conserved,
$N_{{\rm tot}} = N_{+}(t)+N_{-}(t) = a_{+}^{\dagger}(t)a_{+}(t) + 
a_{-}^{\dagger}(t)a_{-}(t) = 2 T_{0}$. The conversion process can 
be studied in terms of the \lq\lq Bloch vector operator"  
${\bf T} = T_{1}{\bf e}_{1} + T_{2}{\bf e}_{2} + T_{3}{\bf e}_{3}$, 
introduced in \cite{Kret1999}, with
\begin{eqnarray}
T_{1}(t) & = & {\textstyle \frac{1}{2}}\left(a_{+}^{\dagger}(t)\,a_{-}(t) + a_{-}^{\dagger}(t)\,a_{+}(t)\right) \,,\\
T_{2}(t) & = & {\textstyle \frac{1}{2i}}\left(a_{+}^{\dagger}(t)\,a_{-}(t) - a_{-}^{\dagger}(t)\,a_{+}(t)\right) \,,\\
T_{3}(t) & = & {\textstyle \frac{1}{2}}\left(a_{+}^{\dagger}(t)\,a_{+}(t) - a_{-}^{\dagger}(t)\,a_{-}(t)\right) \,.
\end{eqnarray}  
The components obey the same commutation rules as an angular momentum, $[T_{j},T_{k}]=i\epsilon_{\!jkl}T_{l}$ 
($j,k,l=1,2,3$) and
$T_{1}^{2}+T_{2}^{2}+T_{3}^{2}=T_{0}(T_{0}+1)$.
The expectation value $\langle {\bf T}\rangle$ is a real 
3-dimensional vector of constant length that describes a 
precessional motion on a \lq\lq Bloch sphere"  during the 
quadrupolar excitation. The model Hamiltonian expressed in terms of 
${\bf T}$ 
\begin{eqnarray}
H(t) & = & \hbar\omega_{1}(T_{0}+{\textstyle \frac{1}{2}}) 
+ \hbar\omega_{c}T_{3}(t)\\
& & +\hbar\cdot 2g\left(\cos\phi_{d}(t)\cdot T_{1}(t) 
+ \sin\phi_{d}(t)\cdot T_{2}(t)\right) \nonumber
\end{eqnarray}
governs the time development of the Bloch vector operator.
Compare this result to the Hamiltonian $H_{{\rm mag}}$ 
that describes the precession of the nuclear spin ${\bf I}$
in magnetic resonance experiments,
${\bf I} = I_{1}{\bf e}_{1} + I_{2}{\bf e}_{2} + I_{3}{\bf e}_{3}$, 
\begin{eqnarray}
H_{{\rm mag}} & = & -\vec{\mu}\cdot{\bf B}\\
& = & -\hbar\omega_{L}I_{3}-\hbar\gamma B_{1}
(\cos\omega t \cdot I_{1}-\sin\omega t \cdot I_{2}),\nonumber
\end{eqnarray}
where $\vec{\mu}=\gamma{\bf I}$ is the nuclear magnetic 
moment, $\gamma$ the gyromagnetic ratio, and $\omega_{L}$ 
the Larmor frequency. Both Hamiltonians govern the time development 
of a vector operator, thus exhibiting a dynamical similarity between 
nuclear magnetic resonance on the one hand and ion motion in a Penning 
trap with quadrupole excitation on the other hand. This structural 
analogy provides deeper insight why Ramsey's idea of using separated 
oscillating fields can be successfully applied also to Penning trap physics.
 
Initially ($t=0$) the ions are prepared in a pure magnetron mode
($N_{{\rm tot}}=N_{-}(0)=-2\langle T_{3}(0)\rangle$), with the 
objective to convert the
magnetron motion as completely as possible into cyclotron motion and
thus to bring the radial energy to its maximum. For the conventional
excitation scheme with a single pulse (see Fig.~\ref{AnregSchema}a)
of duration $\tau$ and constant amplitude the expectation value for
the percentage of converted quanta is obtained as
\begin{equation}
F_1(\omega_R,\tau,g) = \frac{N_{+}(\tau)}{N_{{\rm tot}}} = \frac{4g^2}{\omega^2_R}\cdot\sin^2(\omega_R\tau/2)\mbox{ ,}
\label{F1}
\end{equation}
where $\omega_R=\sqrt{(2g)^2+\delta^2}$ is the analog
of the Rabi frequency and $\delta=\omega_{d}-\omega_{c}$ the 
detuning of the quadrupolar field. The ``conversion time", 
{\it i.e.} the time required for complete conversion 
exactly on resonance, is seen to be $\tau_{c}=\pi/(2g)$. 

If two pulses of quadrupolar radiation, each of duration $\tau_{1}$, 
and separated by
a waiting period $\tau_0$ (see Fig.~\ref{AnregSchema}b), are used
for the excitation, the
expectation value for the percentage of converted quanta becomes
\begin{eqnarray}
& & F_2(\delta,\tau_0,\tau_1,g)=\frac{16g^2}{\omega_R^2} \cdot
 \sin^2\left(\frac{\omega_R\tau_1}{2}\right)
 \\
& &\cdot\left[\cos\frac{\delta\tau_0}{2}\cdot\cos\frac{\omega_R\tau_1}{2}   
 -\frac{\delta}{\omega_R}\cdot\sin\frac{\delta\tau_0}{2} 
   \cdot\sin\frac{\omega_R\tau_1}{2}\right]^2\mbox{.}\nonumber 
\label{F2}
\end{eqnarray}

\begin{figure}
\resizebox{0.4\textwidth}{!}
{%
 \includegraphics{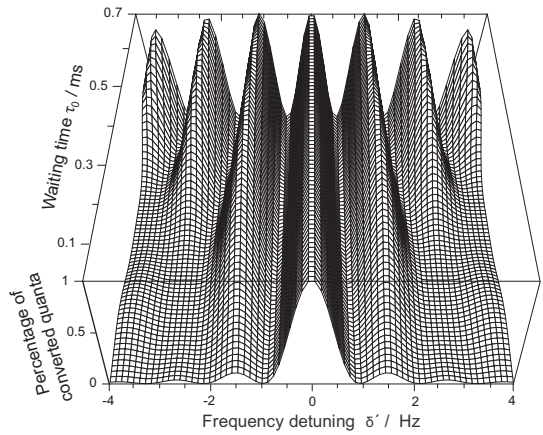}
   } \vspace*{0.0cm}
   \caption{Percentage of converted quanta versus the frequency 
   detuning
   $\delta'=\delta/(2\pi)$ and versus the waiting time $\tau_0$ of 
   the two-pulse Ramsey scheme of Fig.~\ref{AnregSchema}b. Each 
   pulse has a duration of
   $\tau_1=\tau_c/2$ to obtain complete conversion at resonance, 
   the total duration of the excitation
   cycle is fixed to 900 ms.} \label{2PulsEntwicklung3D100}
\end{figure}

\begin{figure}
\resizebox{0.5\textwidth}{!}
{
 \includegraphics{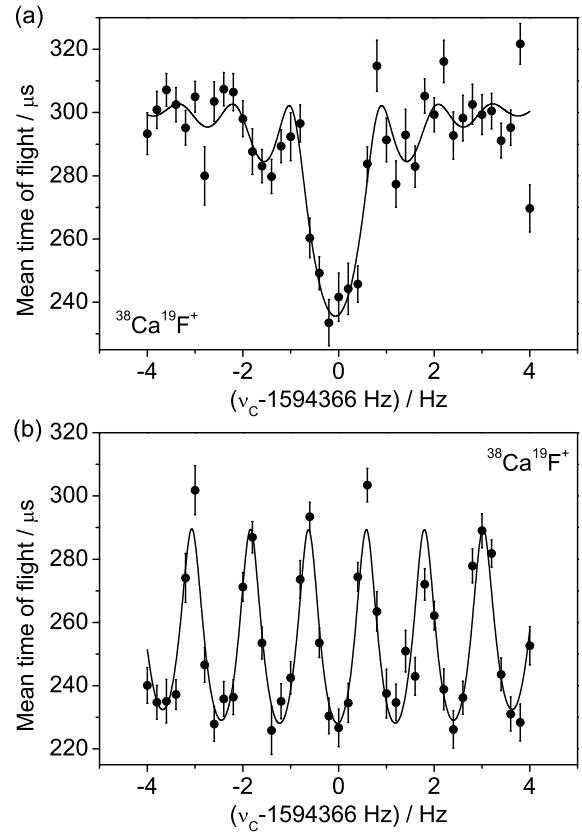}  
        } \vspace*{0.0cm}
   \caption{Time-of-flight ion-cyclotron-resonance spectra of 
   $^{38}\mbox{Ca}^{19}\mbox{F}^+$
   with the conventional quadrupolar excitation (a) and with a 
   two-pulse Ramsey scheme
   with two 100 ms duration excitation periods interrupted by 
   a 700 ms waiting
   period (b). The solid curves are fits of the theoretical 
   line shapes to the data.
   The center frequency, where the detuning is 0, is at 
   1594365.92 Hz.} \label{Ca38F19vertikal}
\end{figure}

In order to compare experimentally the different excitation methods 
the total
duration of the excitation cycle was chosen equal for both, namely
900 ms. The calculated energy conversion is shown in
Fig.~\ref{2PulsEntwicklung3D100} as a
function of the frequency detuning $\delta'=\delta/(2\pi)$ and the 
waiting time $\tau_0$.
The conventional single pulse excitation appears here as the
limiting case with waiting period $\tau_0=0$. The experimental
resonance spectra for a conventional scheme as well as for a scheme 
with two excitation periods of
$\tau_{1}=100$ ms and $\tau_0=700$ ms are shown in
Fig.~\ref{Ca38F19vertikal}. In the latter case the sidebands are very 
pronounced and
the full-width-at-half-maximum (FWHM) is considerably reduced. The solid 
line represents the fit of the theoretical line shape to the data points. 

 
The mass measurements on the short-lived nuclide $^{38}\mbox{Ca}$ 
using the Ramsey method
were performed using the Penning trap mass spectrometer
ISOLTRAP~\cite{Blau2005} installed at the on-line mass 
separator
facility ISOLDE/CERN. The calcium isotope was
produced by bombarding a heated titanium metal foil 
target with
1.4-GeV protons from the CERN proton-synchrotron-booster
accelerator. A hot tungsten surface was used to ionize 
the released
atoms. In order to suppress isobaric contaminations by 
$^{38}\mbox{K}^+$ ions, a
CF$_4$ leak was added and the ions of interest were delivered
to ISOLTRAP in form of the molecular sideband
$^{38}\mbox{Ca}^{19}\mbox{F}^+$. Ions extracted from 
the source were
accelerated to 30~keV, mass separated in ISOLDE's 
high-resolution
mass separator, and injected into the first part of the
ISOLTRAP apparatus, a gas-filled linear radiofrequency 
quadrupole
ion trap for accumulation, cooling and bunching of the ion
beam~\cite{Herf2001}. From here the ions were transferred at lower
energy as short bunches to a cylindrical Penning trap for further
buffer-gas cooling and isobaric purification~\cite{Sava1991}. 
The actual mass measurement was
performed in a second, hyperboloidal Penning trap~\cite{Boll1996}
via the cyclotron frequency determination.

The time-of-flight ion
cyclotron resonance detection technique~\cite{Grae1980} relies 
on the coupling of the
ion's orbital magnetic moment to the magnetic field gradient after
excitation of the ion motion with rf-fields and axial ejection 
from the trap into a time-of-flight section. The sum frequency 
of the modified
cyclotron mode and the magnetron mode $\nu_c=\nu_++\nu_-$ is probed
by the measurement of the radial energy~\cite{Boll1990,Blau2003b}. 
A typical
time-of-flight cyclotron resonance of $^{38}\mbox{Ca}^{19}\mbox{F}^+$
ions using the conventional excitation technique with one continuous
rf pulse (Fig.~\ref{AnregSchema}a) is shown in
Fig.~\ref{Ca38F19vertikal}a. The mass of the ion of interest is
obtained from a comparison of its cyclotron frequency with the one
of a reference ion with well-known mass, here $^{39}\mbox{K}^+$.

The uncertainty of the center frequency depends mainly on three
parameters. First, the relative uncertainty is inversely proportional
to the square root of the number of recorded ions. Second, it is
proportional to the linewidth of the resonance and third it depends
on the shape of the resonance and its sidebands. The stronger the
sidebands are, the more precisely the center frequency can be
determined. For the latter two aspects Ramsey's excitation method
with separated oscillatory fields is superior to the conventional method.
\begin{figure}
\resizebox{0.5\textwidth}{!}
{
 \includegraphics{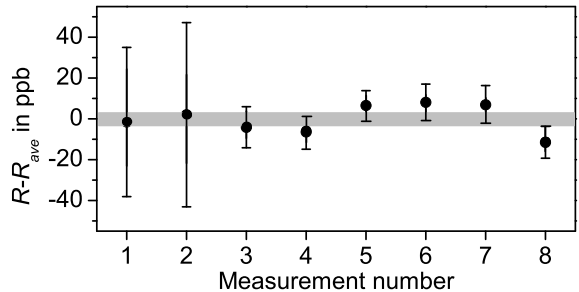}
   } \vspace*{0.0cm}
   \caption{Difference of the measured cyclotron frequency ratios $R$ and 
   their average value
   $R_{ave}$ between $^{38}\mbox{Ca}^{19}\mbox{F}^+$ and $^{39}\mbox{K}^+$. 
   The error bars are
   the statistical uncertainties. Each data point contains the same number 
   of recorded ions in the resonance.}
\label{Ratio}
\end{figure}

Altogether eight resonances of $^{38}\mbox{Ca}^{19}\mbox{F}^+$, each with a 
scan width of $\pm4$ Hz around the expected cyclotron frequency were
measured with the Penning trap mass spectrometer ISOLTRAP. The first two 
employed the conventional
excitation scheme with 900 ms,
the others used the  Ramsey-type scheme of two
100-ms excitation pulses and one 700-ms waiting period (
Fig~\ref{AnregSchema}). Thus, the
total excitation period of both schemes was 900 ms and they can be
easily compared. Taking only into account the statistical
error, the mass excess can be determined to within a
1.03\,keV uncertainty for the two resonances with conventional 
excitation. The
uncertainty of the mass excess obtained from two Ramsey type
measurements under identical experimental conditions is here
drastically reduced to 0.25\,keV. All six measurements lead to an
overall statistical uncertainty of the mass excess of only
0.12\,keV. After consideration of
systematic uncertainties of ISOLTRAP~\cite{Kell2003} we
obtain finally a frequency ratio $R$ between the ion of interest 
and the reference ion 
$\nu_{{\rm ref}}/\nu_{{\rm ion}}=1.462\,257\,6087(50)(229)$,
where the first error derives from the statistics and the second 
one denotes the systematic error of the measurement. Together 
with the
mass excess of 
$^{19}F$ $ME$($^{19}\mbox{F}$)=-1487.39(07)\,keV~\cite{Audi2003} 
we obtain a mass excess $ME$($^{38}\mbox{Ca}$)=-22058.01(65)\,keV,
which is in agreement with the previously accepted
value~\cite{Audi2003} as well as with the recent value
$ME$($^{38}\mbox{Ca}$)=-22058.53(28)\,keV~\cite{Boll2006}, measured 
by the new Penning trap facility at MSU and being a factor of two 
more precise.  

The 12 best known superallowed Fermi-type beta decays ranging from
$^{10}\mbox{C}$ to $^{74}\mbox{Rb}$ yield an average Ft value of
$\overline{Ft}=3072.7(0.8)\mbox{s}$~\cite{Hard2005b}, which
confirms the CVC hypothesis at a level of $3\cdot 10^{-4}$. Further
mass measurements on superallowed $\beta$-emitters for an even more
stringent CVC test have been performed recently
\cite{Sava2005,Eron2006,Boll2006}, demonstrating the continuous
interest in this weak interaction and Standard Model test.
$^{38}\mbox{Ca}$ is a new candidate for testing the CVC hypothesis. 
Its large isospin symmetry-breaking correction of
0.73(5)\% makes it a proper candidate for testing the validity of
the calculations. Until recently the Ft value had a relative
uncertainty of $4\cdot 10^{-3}$, limited by the uncertainty of the
ground-state mass, the uncertainty of the $0^+\rightarrow0^+$ branching
ratio, and the accuracy of the half-life $(T_{1/2}=440(8) ms)$,
which was measured and improved in parallel to our experiment at
ISOLDE. Here, the data analysis is presently under way.

Obviously, our accuracy is not limited by the statistical uncertainty, 
but by the systematic error, which comes from magnetic field 
fluctuations and mass dependent error of the experimental setup. 
In the future these systematic errors can be reduced by a 
stabilization of the magnetic field and a local determination 
of mass dependent deviations with carbon clusters~\cite{Kell2003}.

In conclusion, the method of separated oscillatory fields
has been for the first time theoretically derived and applied 
to the mass measurement of
radioactive ions confined in a Penning trap.  It
allows a considerable reduction in data-taking time by its
improvement of precision with much less statistics. Thus, it 
is ideally suited for radionuclides,
especially of half-lives below 100\,ms and
production rates of only a few 100 ions/s where the uncertainty 
is often
limited by the statistical error, as in the case of
$^{32}$Ar $(T_{1/2}=98 ms)$~\cite{Blau2003} and 
$^{74}$Rb $(T_{1/2}=65 ms)$~\cite{Kell2004}.

This work was supported by the German Ministry for Education 
and Research (BMBF) under contracts 06GF151 and 06MZ215, by the 
European Commission under contracts HPMT-CT-2000-00197 
(Marie Curie Fellowship) and RII3-CT-2004-506065 (TRAPSPEC), 
and by the Helmholtz association of national research centers 
(HGF) under contract VH-NG-037. We also thank the ISOLDE 
Collaboration as well as the ISOLDE technical group for their assistance.


\begin{thebibliography}{99}
\bibitem{Rams1990} N.F. Ramsey, Rev. Mod. Phys. \textbf{62}, 541 (1990).
\bibitem{Rams1949} N.F. Ramsey, Phys. Rev. \textbf{76}, 996 (1949).
\bibitem{Rams1950} N.F. Ramsey, Phys. Rev. \textbf{78}, 695 (1950).
\bibitem{Boll1992} G. Bollen \emph{et al.}, Nucl. Instrum. Meth. B \textbf{70}, 490 (1992).
\bibitem{Berg2002} I. Bergstr\"om \emph{et al.}, Nucl. Instrum. Meth. A \textbf{487}, 618 (2002).
\bibitem{Hard2005a} J.C. Hardy and I.S. Towner, Phys. Rev. C \textbf{71}, 055501 (2005).
\bibitem{Hard2005b} J.C. Hardy and I.S. Towner, Phys. Rev. Lett. \textbf{94}, 092502 (2005).
\bibitem{Grae1980} G. Gr\"aff \emph{et al.}, Z. Phys. A \textbf{297}, 35 (1980).
\bibitem{Koen1995} M. K\"onig \emph{et al.}, Int. J. Mass Spectrom. \textbf{142}, 116 (1995).
\bibitem{Blau2006} K. Blaum, Phys. Rep. \textbf{425}, 1 (2006).
\bibitem{Brow1986} L.S. Brown and G. Gabrielse, Rev. Mod. Phys. \textbf{58}, 233 (1986).
\bibitem{Kret1999} M. Kretzschmar, in {\it Trapped Charged Particles and Fundamental Physics},
ed. by Daniel H. E. Dubin and Dieter Schneider,
AIP Conf. Proc. Vol. 457, p. 242 (1999).
\bibitem{Blau2005} K. Blaum \textit{et al.}, Nucl. Phys. A \textbf{752}, 317 (2005).
\bibitem{Herf2001} F. Herfurth \emph{et al.}, Nucl. Instr. Meth. A \textbf{469}, 254 (2001).
\bibitem{Sava1991} G. Savard \emph{et al.}, Phys. Lett. A \textbf{158}, 247 (1991).
\bibitem{Boll1996} G. Bollen \emph{et al.}, Nucl. Instr. and Meth \textbf{368}, 675 (1996).
\bibitem{Blau2003b} K. Blaum \textit{et al.}, J. Phys. B \textbf{36}, 921 (2003).
\bibitem{Boll1990} G. Bollen \emph{et al.}, J. Appl. Phys. \textbf{68}, 4355 (1990).
\bibitem{Kell2003} A. Kellerbauer \emph{et al.}, Eur. Phys. J. D \textbf{22}, 53 (2003).
\bibitem{Audi2003} G. Audi \emph{et al.}, Nucl. Phys. A \textbf{729}, 337 (2003).
\bibitem{Boll2006} G. Bollen \emph{et al.}, Phys. Rev. Lett. \textbf{96}, 152501 (2006).
\bibitem{Sava2005} G. Savard \emph{et al.}, Phys. Rev. Lett. \textbf{95}, 102501 (2005).
\bibitem{Eron2006} T. Eronen \emph{et al.}, Phys. Lett. B \textbf{636}, 191 (2006).
\bibitem{Blau2003} K. Blaum \emph{et al.}, Phys. Rev. Lett. \textbf{91}, 260801 (2003).
\bibitem{Kell2004} A. Kellerbauer \emph{et al.}, Phys. Rev. Lett. \textbf{93}, 072502 (2004).

\end{thebibliography}
\end{document}